# A Decade of NASA Strategic Astrophysics Technology Investments: Technology Maturation, Infusion, and Other Benefits

Thai Pham[1], Opher Ganel, and Azita Valinia[2]
Physics of the Cosmos and Cosmic Origins Program Office, NASA Goddard Space Flight Center,
8800 Greenbelt Rd., Greenbelt, MD 20771, USA

Nicholas Siegler and Brendan Crill
Exoplanet Exploration Program Office, Jet Propulsion Laboratory, 4800 Oak Grove Dr., Pasadena, CA 91109, USA

Mario R. Perez
Astrophysics Division, NASA Headquarters, Washington DC 20546, USA

## ABSTRACT

NASA's Astrophysics Division (APD) funds development of cutting-edge technology to enable its missions to achieve ambitious and groundbreaking science goals. These technology development efforts are managed by the Physics of the Cosmos (PCOS), Cosmic Origins (COR), and Exoplanet Exploration (ExE) Programs. The NASA Strategic Astrophysics Technology (SAT) Program was established in 2009 as a new technology maturation program to fill the gap in the Technology Readiness Level (TRL) range from 3 to 6. Since program inception, 100 SAT grants have been openly competed and awarded, along with dozens of direct-funded projects, leading to a host of technologies advancing their TRLs and/or being infused into space and suborbital missions and ground-based projects. We present the portfolio distribution in terms of specific technology areas addressed, including optics, detectors, coatings, coronagraphs, starshades, lasers, electronics, cooling systems, and micro-thruster subsystems. We show an analysis of the rate of TRL advances, infusion success stories, and other benefits such as training the future astrophysics workforce, including students and postdoctoral fellows hired by projects. Finally, we present APD's current strategic technology maturation priorities for investment, enabling a range of future strategic astrophysics missions.

## 1. INTRODUCTION

Since its inception, NASA has participated in both space exploration and the scientific quest to discover the origin of the universe and expand our understanding of how it works. Within the realm of space sciences, astrophysics represents humanity's yearning to discover distant worlds and the first events in the history of the universe. These observational efforts inescapably lead to new questions that motivate new measurements and new missions [1]. However, the main obstacle to reaching the next level of detection in observational astrophysics is insufficient technology readiness of essential components required to enable these new missions. Furthermore, astrophysics is often a photon-starved discipline, demanding exquisite performance from all systems and subsystems used in on-sky observation and detection. The challenges are many and the path forward, while demanding, is also exciting.

Within the Science Mission Directorate (SMD) at NASA Headquarters, APD is responsible for studying the universe and our place within it. To this end, APD established three science-themed Programs that cover three of the most fundamental questions posed by humanity.

- "*How does the universe work?*" is covered by the PCOS Program.
- "*How did we get here?*" is addressed by the COR Program.
- "*Are we alone?*" is the focus of the ExE Program (ExEP).

These enduring questions have been under study in one form or another for millennia, with each generation delving deeper, as our tools and understanding improved. Recognizing that technology readiness is now one of the most important limiting factors on progress in observational astronomy and astrophysics, APD established three programs that fund technology development and maturation: (1) SAT, (2) Astrophysics Research and Analysis (APRA), and (3) Roman Technology

---

[1] Thai.Pham@nasa.gov; phone: +1-301-286-4809
[2] Currently Chief Scientist, NASA Engineering & Safety Center (NESC)

Fellowship (RTF). All three are openly solicited through NASA's omnibus Research Opportunities in Earth and Space Science (ROSES) announcement of opportunity (AO) [2].

The SAT program, established in 2009, is a technology maturation program filling the so-called "mid-TRL gap" between TRL 3 and 6 for technologies that enable or enhance future strategic astrophysics missions.

APRA funds technology development across the full TRL spectrum, from 1 to 9 (as well as scientific investigations unrelated to technology development) and is not necessarily limited to strategic missions.

RTF targets early-career researchers to develop the skills needed to lead astrophysics technology development projects and future astrophysics missions, develop innovative strategic technologies, and pursue long-term positions.

Anticipating the need for future large space telescopes, in 2018 APD started the System-Level Segmented-Telescope Design (SLSTD) program. This openly competed program funds industry to develop architectures and technologies that would enable large, highly stable, segmented-aperture space telescopes.

Strategic astrophysics missions are agency-led missions or concepts that APD is developing, participating in, or interested in, to respond to high-priority science questions. This could be triggered by their being prioritized by the current Decadal Survey; being identified for implementation in the Astrophysics Implementation Plan; and/or inspiring broad community interest such as captured in the visionary 30-year Astrophysics Roadmap, Enduring Quests, Daring Visions [3].

In addition, APD also direct-funds technology development projects through e.g. the Internal Scientist Funding Model (ISFM) that funds core scientific efforts involving NASA scientists. Additional directed investment is managed by study offices responsible for developing potential US technology contributions to two European Space Agency (ESA) missions, the Advanced Telescope for High-ENergy Astrophysics (ATHENA) [4] and the Laser Interferometer Space Antenna (LISA) [5]. US participation in these missions is also considered strategic by APD.

The currently funded strategic astrophysics missions in formulation or implementation are:

- The James Webb Space Telescope (JWST) [6]
- The Roman Space Telescope (formerly the Wide Field InfraRed Survey Telescope, WFIRST) [7]
- ESA's Euclid mission [8]
- The Japan Aerospace Exploration Agency (JAXA) X-Ray Imaging and Spectroscopy Mission (XRISM) [9]

As mentioned above, ESA missions identified for potential US contributions are:

- ATHENA
- LISA

The large-mission concept studies submitted to the 2020 Astronomy and Astrophysics Decadal Survey panel are:

- The Habitable Exoplanet Observatory (HabEx) [10]
- The Large UV/Optical/IR Surveyor (LUVOIR) [11]
- The Lynx X-ray observatory (formerly X-ray Surveyor) [12]
- Origins (formerly Origins Space Telescope or Far-IR Surveyor) [13]

A currently operating strategic platform with ongoing technology needs is:

- The Stratospheric Observatory for Infrared Astronomy (SOFIA) [14]

Other strategic mission concepts per the 2010 Astronomy and Astrophysics Decadal Survey report [15] and/or the Astrophysics Roadmap are:

- The Cosmic Microwave Background (CMB) Polarization Surveyor, Inflation Probe, or Probe of Inflation and Cosmic Origins (PICO) [16]
- The Black-Hole Mapper
- The Cosmic-Dawn Mapper
- The Exo-Earth Mapper
- The Gravitational-Wave Mapper

Technology development for funded missions such as the above four in formulation or implementation becomes the responsibility of those missions. Technology development for ATHENA and LISA is managed by their two mission-specific study offices. Technology development relevant to the four large-mission concepts, SOFIA, and Inflation Probe is funded through SAT, ISFM, SLSTD, APRA, and certain direct-funded programs.

Maturing enabling and enhancing technologies for strategic missions is challenging, given the extreme requirements of these visionary missions. For example, LISA must measure the 2.5-million-km distance between each pair of the three satellites in the constellation to picometer precision – less than 2% the size of a hydrogen atom. LUVOIR, as another example, would require maintaining a 15-m-diameter segmented telescope stable to tens of picometers over a 10-minute observation. The optics of Lynx would need to bend X-ray photons' trajectories through grazing-incidence mirrors with a precision of 0.5 arcsec – 1/3600 of 1 degree. HabEx and LUVOIR need to block out the light arriving from a target star while observing the far dimmer light reflecting off a planet orbiting that star, requiring a contrast ratio of $10^{10}$ at visible wavelengths. This would be the equivalent of being able to see the light output of a single candle right next to a spotlight over 9000 times brighter than the brightest spotlight in the world.

Since 2017, the three APD Programs (PCOS, COR, and ExEP) have closely collaborated and coordinated their technology development management efforts. This includes using a unified technology-gap solicitation form; unified gap-prioritization process, criteria, and metrics; participation of technologists from each of the Programs in the others' prioritization discussions; and merging the three resulting gap priority lists into a unified APD technology-gap priority list.

Prioritization was carried out annually until 2017, at which point it was changed to a biennial cadence to better match the longer technology development lifecycles. The most recent prioritization was published in 2019. The next prioritization is scheduled for late 2021, following the planned release of the 2020 Astronomy and Astrophysics Decadal Survey.

## 2. PROGRAM STATUS

APD has been preparing for the 2020 Decadal Survey for several years, with Science and Technology Definition Teams (STDTs) established in early 2016 for each of four large-mission concepts, HabEx, LUVOIR, Lynx, and Origins. The STDTs were each chartered to develop a compelling science case, a design reference mission (DRM) to achieve the science goals, and a technology roadmap to enable the DRM. The STDTs each submitted several deliverables, culminating with their final reports which APD forwarded to the 2020 Astronomy and Astrophysics Decadal Survey panel. The three APD Programs folded technology gaps submitted by each of the STDTs into their gap prioritization process in 2016, 2017, and 2019 (APD stopped the annual process in 2017 and began a biennial process as of 2019). This included the needs of the four mission concept studies' reference designs in APD's prioritized technology gaps. APD also funded 10 Probe mission studies, which assume mission cost caps of $1 billion. These studies mostly assumed existing technologies, though some would benefit from technology advances funded by APD.

The 2019 prioritized Astrophysics technology gaps list is presented in Section 4. That list (as did its earlier versions), along with programmatic and other considerations, informs APD on which technologies to solicit, and which technology development proposals to fund. The current portfolio of technology maturation investments includes 59 active projects managed by the three Programs.

The PCOS Program focuses primarily on technologies enabling measurement of gravitational waves, microwaves, and X rays, in pursuit of understanding some of science's most profound phenomena. This includes testing the validity of Einstein's General Theory of Relativity; and understanding the nature of spacetime, the behavior of matter and energy in extreme environments, the cosmological parameters governing inflation and the evolution of the universe, and the nature of dark matter and dark energy. The 25 technology investigations in the current PCOS Program portfolio are shown in Table 1.

| Title | PI | PI Org | Tech Type |
|---|---|---|---|
| Magnetically Coupled Calorimeters | Bandler, Simon | GSFC | Detector |
| Toward Fast, Low-Noise, Radiation-Tolerant X-ray Imaging Arrays for Lynx: Raising Technology Readiness Further | Bautz, Mark | MIT | Detector |
| Microwave SQUID readout technology to enable Lynx and other future Great Observatories | Bennett, Douglas | NIST | Electronics |
| Direct Fabrication of X-Ray Mirror Full Shells | Bongiorno, Stephen | MSFC | Optics |
| Low Stress X-Ray Mirror Coatings | Broadway, David | MSFC | Optical Coating |

| Title | PI | PI Org | Tech Type |
|---|---|---|---|
| Hybrid X-ray Optics by Additive Manufacturing | Broadway, David | MSFC | Optics |
| UV LED-based Charge Management System | Conklin, John | U Florida | Electronics |
| Computer-Controlled Polishing of High-Quality X-Ray Mirror Mandrels | Davis, Jacqueline | MSFC | Optics |
| Microwave Multiplexing Readout Development | Frisch, Josef | SLAC | Electronics |
| Differential Deposition for Figure Correction in X-ray Optics | Kilaru, Kiran | MSFC | Optics |
| TES Microcalorimeters | Kilbourne, Caroline | GSFC | Detector |
| Providing Enabling and Enhancing Technologies for a Demonstration Model of the Athena X-IFU | Kilbourne, Caroline | GSFC | Detector |
| Phase Measurement System Development for Interferometric Gravitational Wave Detectors | Klipstein, William | JPL | Electronics |
| Advancing the Focal Plane TRL for LiteBIRD and Other Next Generation CMB Space Missions | Lee, Adrian | UC Berkeley | Detector |
| Telescopes for Space-Based Gravitational-Wave Observatories | Livas, Jeffrey | GSFC | Telescope |
| Development of Low Power FPGA-based Readout Electronics for Superconducting Detector Arrays | Mauskopf, Philip | ASU | Electronics |
| Superconducting Antenna-Coupled Detectors and Readouts for CMB Polarimetry in PICO | O'Brient, Roger | JPL | Detector |
| Laboratory Spectroscopy for Space Atomic Physics | Porter, Scott | GSFC | Detector |
| X-ray Testing and Calibration | Ramsey, Brian | MSFC | Optics |
| Development of Adjustable X-ray Optics with 0.5 Arcsecond Resolution for the Lynx Mission Concept | Reid, Paul | SAO | Optics |
| High Resolution and High Efficiency X-ray Transmission Grating Spectrometer | Schattenburg, Mark | MIT | Optics |
| Readying X-ray Gratings and Optics for Space Applications; Manufacturability and Alignment | Smith, Randall | SAO | Optics |
| Space-based Gravitational Wave Laser Technology Development Project for the LISA Mission | Yu, Anthony | GSFC | Laser |
| Next Generation X-ray Optics | Zhang, William | GSFC | Optics |
| LISA Colloid Microthruster Technology | Ziemer, John | JPL | Micro-propulsion |

**Table 1.** Current PCOS Program strategic technology portfolio (PI, Principal Investigator; GSFC, Goddard Space Flight Center; MIT, Massachusetts Institute of Technology; SQUID, Superconducting QUantum Interference Device; NIST, National Institute of Standards and Technology; MSFC, Marshall Space Flight Center; SLAC, Stanford Linear Accelerator Laboratory; TES, Transition-Edge Sensor; X-IFU, X-ray Integral Field Unit; JPL, Jet Propulsion Laboratory; UC Berkeley, University of California, Berkeley; FPGA, Field-Programmable Gate Array; ASU, Arizona State University; SAO, Smithsonian Astrophysical Observatory).

The COR Program focuses primarily on technologies for measuring UV, visible light, and IR in pursuit of understanding when the first stars in the universe formed and how they influenced the environments around them; how the pervasive and mysterious dark matter clumped up early in the life of the universe, pulling gas along with it into dense concentrations that eventually became galaxies; how galaxies evolved from the very first systems to the types we observe "in the here and now," such as our Milky Way; and understanding when in the early universe supermassive black holes first formed and how they have affected the galaxies in which they reside. The 17 technology investigations in the current COR Program portfolio are shown in Table 2.

| Title | PI | PI Org | Tech Type |
|---|---|---|---|
| Photon Counting NIR LmAPD Arrays for Ultra-low Background Space Observations | Bottom, Michael | U Hawaii, Honolulu | Detector |
| Ultrasensitive Bolometers for Far-IR Spectroscopy at the Background Limit | Bradford, Charles | JPL | Detector |
| Ultra-Stable Telescope Research and Analysis -Technology Maturation (ULTRA-TM) | Coyle, Laura | Ball Aerospace | Telescope |
| A Single Photon Sensing and Photon Number Resolving Detector for NASA Missions | Figer, Donald | RIT | Detector |

| Title | PI | PI Org | Tech Type |
|---|---|---|---|
| Electron Beam Lithography Ruled Gratings for Future UV/Optical Missions: High-Efficiency and Low-Scatter in the Vacuum UV | Fleming, Brian | U Colorado | Optics |
| Scalable micro-shutter systems for UV, visible, and IR spectroscopy | Greenhouse, Matthew | GSFC | Optics |
| High Performance, Stable, Scalable UV Aluminum Mirror Coatings Using ALD | Hennessy, John | JPL | Optical Coating |
| Development of High-Resolution Far-IR Array Receivers | Mehdi, Imran | JPL | Detector |
| Development of Digital Micromirror Devices (DMD) for Far-UV Applications | Ninkov, Zoran | RIT | Optics |
| Technology Maturation for Astrophysics Space Telescopes (TechMAST) | Nordt, Alison | Lockheed Martin | Telescope |
| Electron-beam Generated Plasma to Enhance Performance of Protected Aluminum Mirrors for Large Space Telescopes | Quijada, Manuel | GSFC | Optical Coating |
| Ultra-Stable Structures Development and Characterization Using Spatial Dynamic Metrology | Saif, Babak | GSFC | Metrology/Structure |
| High Performance Sealed Tube Cross Strip Photon Counting Sensors for UV-Vis Astrophysics Instruments | Siegmund, Oswald | UC Berkeley | Detector |
| Development of a Robust, Efficient Process to Produce Scalable, Superconducting kilopixel Far-IR Detector Arrays | Staguhn, Johannes | JHU | Detector |
| Predictive Thermal Control (PTC) Performance Tests | Stahl, H. Philip | MSFC | Optics |
| High-Efficiency Continuous Cooling for Cryogenic Instruments and sub-Kelvin Detectors | Tuttle, James | GSFC | Cooling System |
| Large Format, High Dynamic Range UV Detector Using MCPs and Timepix4 Readouts | Vallerga, John | UC Berkeley | Detector |

**Table 2.** Current COR Program strategic technology portfolio (LmAPD, Linear-mode Avalanche PhotoDiode; RIT, Rochester Institute of Technology; ALD, Atomic Layer Deposition; MCP, Micro-Channel Plate).

The ExEP focuses primarily on technologies that enable the detection and characterization of planets around nearby Sun-like stars, especially Earth-like planets in the habitable zones of Sun-like stars; and searching for signatures of life. At this time, technology needs include ultra-stable space-telescope architectures, starshades, coronagraphs, optics, and detectors enabling direct imaging and characterization of exo-Earths. Other current investments include technologies for measurement of planet mass through stellar reflex motion, and for characterizing rocky planets orbiting M-dwarfs through transit spectroscopy.

The 17 technology investigations in the current ExEP portfolio are shown in Table 3.

| Title | PI | PI Org | Tech Type |
|---|---|---|---|
| Laboratory Demonstration of High Contrast Using Phase-Induced Amplitude Apodization Complex Mask Coronagraph (PIAACMC) on a Segmented Aperture | Belikov, Ruslan | ARC | Coronagraph |
| Laboratory Demonstration of Multi-Star Wavefront Control in Vacuum | Belikov, Ruslan | ARC | Coronagraph |
| Development of a Method for Exoplanet Imaging in Multi-Star Systems | Belikov, Ruslan | ARC | Coronagraph |
| MEMS Deformable Mirror Technology Development for Space-Based Exoplanet Detection | Bierden, Paul | BMC | Coronagraph |
| Segmented Coronagraph Design and Analysis study | Chen, Pin | JPL | Coronagraph |
| Linear Wavefront Control for High Contrast Imaging | Guyon, Olivier | U Arizona | Coronagraph |
| A Novel Optical Etalon for Precision Radial Velocity Measurements | Leifer, Stephanie | JPL | EPRV |
| Optimal Spectrograph and Wavefront Control Architectures for High-Contrast Exoplanet Characterization | Mawet, Dimitri | Caltech | Coronagraph |
| Vortex Coronagraph High Contrast Demonstrations | Serabyn, Eugene | JPL | Coronagraph |
| Broadband Light Rejection with the Optical Vortex Coronagraph | Serabyn, Eugene | JPL | Coronagraph |
| First System-level Demonstration of High-Contrast for Future Segmented Space Telescopes | Soummer, Rémi | STSci | Coronagraph |
| Development of an Ultra-Stable Mid-Infrared Detector Array for Space-Based Exoplanet Transit Spectroscopy | Staguhn, Johannes | JHU | Detector |
| Super Lyot ExoEarth Coronagraph | Trauger, John | JPL | Coronagraph |
| Starshade Starlight Suppression | Willems, Phil | JPL | Starshade |
| Starshade Large-Structure Precision Deployment and Stability | Willems, Phil | JPL | Starshade |
| Radiation Tolerant, Photon Counting, Visible and Near-IR Detectors for Space Coronagraphs and Starshades | Rauscher, Bernard | GSFC | Detector |
| Environmental Testing of MEMS Deformable Mirrors | Mejia Prada, Camilo | JPL | Coronagraph |

**Table 3.** Current ExEP Program strategic technology portfolio (ARC, Ames Research Center; BMC, Boston Micromachines Corporation; EPRV, Extreme Precision Radial Velocity; Caltech, California Institute of Technology; STSci, Space Telescope Science Institute; JHU, Johns Hopkins University).

Information on active and completed technology investigations is available via a searchable Astrophysics database or NASA's TechPort. The Programs are currently adding information about hundreds of APRA and RTF awards.

### 3. METRICS AND RESULTS

The analysis presented below covers the 10 years since the SAT Program was initiated, during which APD has funded 147 technology investigations including ISFMs and other projects. To analyze the impacts of this technology maturation investment, 27 of the 147 were removed from our statistics for the following reasons:

- Nine small ISFM efforts or multi-part investigations, too small to be considered strategic investments
- Two architecture studies
- 14 too new to have had time to achieve results
- Two investigations working on topics other than space technologies

Since fully maturing strategic technologies is not typically expected within the schedule and/or budget of a single SAT award, many PIs receive continued funding (competed or directed). Since the original and follow-on investigations have the same technology objective, we combined them for the purpose of the analysis presented here. Thus, the 120 retained investigations were combined into 70 distinct projects, with 30 of these having one or more additional funding cycles, up to a total of five, ranging from 1 to 11 years. The average project duration was 4.0 years (4.1 years for COR, 4.0 for PCOS and 3.7 for ExEP).

Figure 1 shows the distribution of projects between the three Programs, with 23 in COR, 22 in ExEP, and 25 in PCOS. The total investment to date was $209M, of which $60M (29%) in COR, $47M (22%) in ExEP, and $102M (49%) in PCOS, with an average investment of $3.0M per project ($2.6M for COR, $2.1M for ExEP, and $4.1M for PCOS).

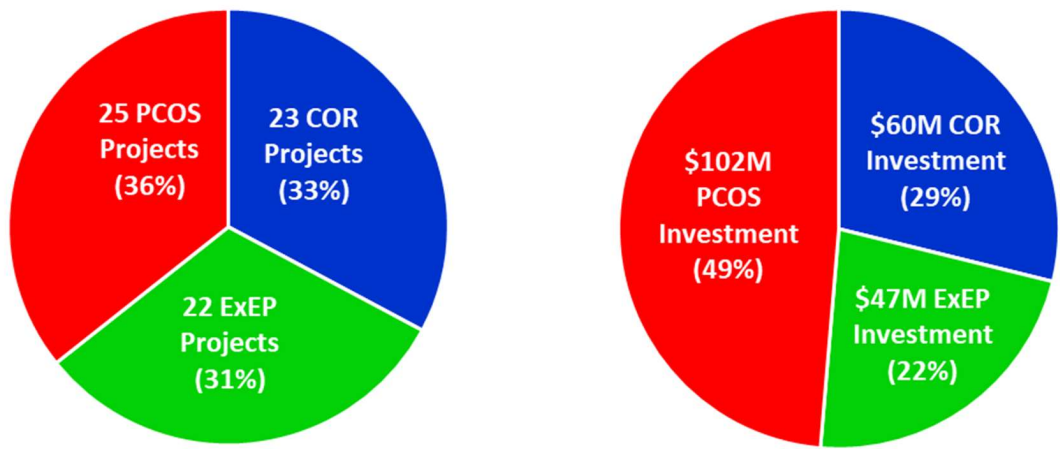

**Fig. 1**. Strategic technology development by Program since 2009: number of distinct projects (left) and total investment (right).

The distribution of organizational types receiving awards to date is shown in Fig. 2. Government labs include Federally Funded Research and Development Centers (FFRDCs) such as JPL and NIST.

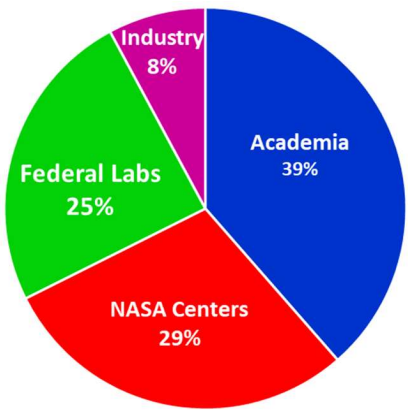

**Fig. 2.** Distribution institutional types receiving awards for strategic technology developments.

The SAT Program investments are distributed among a variety of technology development areas, including detectors, coronagraphs, optics, electronics, optical coatings, starshades, lasers, micropropulsion, cooling systems, picometer-level metrology, and telescopes. The distribution of the technology portfolio is summarized in Fig. 3. Unsurprisingly, detectors, coronagraphs, and optics dominate the portfolio at a combined 67%, given the important roles these technologies play in the portfolio of strategic Astrophysics missions.

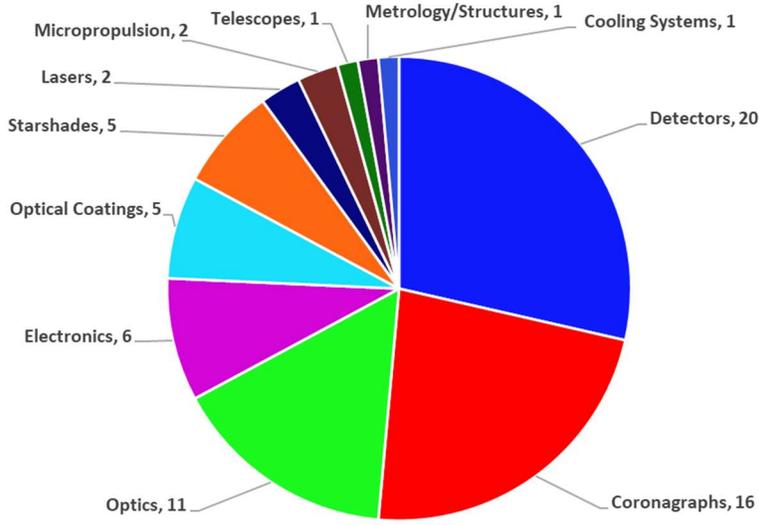

**Fig. 3**. Distribution of the 70 projects in the strategic technology portfolio by topic since SAT inception.

For the following discussion, we reference NASA TRL definitions as explained in the NASA Systems Engineering Processes and Requirements [17].

Few if any technology development projects are expected to mature their technologies all the way to TRL 6 during a grant's period of performance. TRL 6 relates particularly to an integrated system specific to a defined mission or application, which is beyond the normal scope of SAT grants. Although all projects are expected to make progress toward the next TRL, not all are expected to advance by even one level. This is because advancing strategic technologies, by its nature, requires moving beyond the cutting edge of what is currently possible. In addition, where an investigation addresses a system-level problem, the TRL is limited by the lowest-TRL subsystem, so even if the TRL of one or more subsystems advances, if the lowest-TRL subsystem or component remains at its original level, so does the overall system. Next, the TRL scale is a step-function, and if even one requirement for achieving the next level isn't met, the TRL cannot advance. Worse yet, in some cases, the target application of a technology changes over time, making the TRL advance a moving target. For example, one investigation initially worked to mature X-ray mirror technology for the International X-ray Observatory, then changed to the ATHENA mission, and then to the Lynx mission concept. This evolution tightened the angular resolution requirement tenfold, from 5 arcsec to 0.5 arcsec.

Overall, 37% of strategic technology projects (26 of 70) advanced by at least one TRL, of which 9% (six) advanced by two (Fig. 4). The high-risk, high-reward nature of the strategic technology development portfolio makes this result very impressive. It is worth noting that at least half the projects in the key areas of optics, starshades, electronics, lasers, micropropulsion, and telescopes advanced their TRL by at least one level.

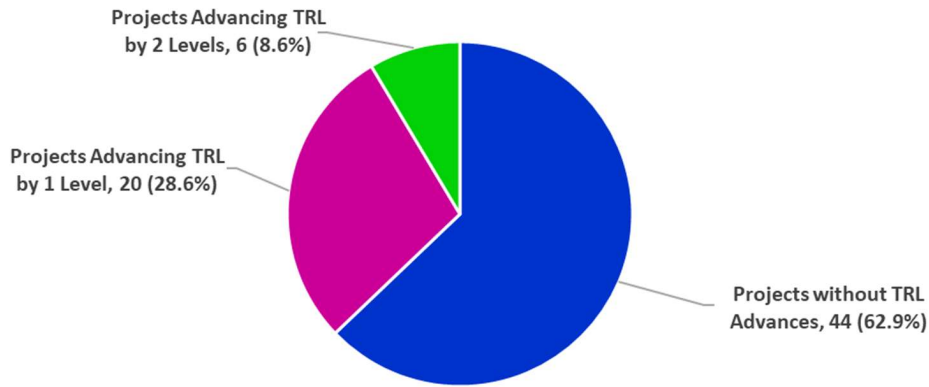

**Fig. 4.** Distribution of TRL advancement while in the program for strategic technology development projects.

We observe that the higher an investigation's entry TRL, the more difficult it was to complete all the requirements for advancing to the next TRL (Fig. 5). While the statistical sample is small, especially for investigation entry TRL ($TRL_{in}$) of 2 and 5, the trend is clear and unsurprising. Counting advances by two TRLs twice, the portfolio had a total of 32 TRL advances.

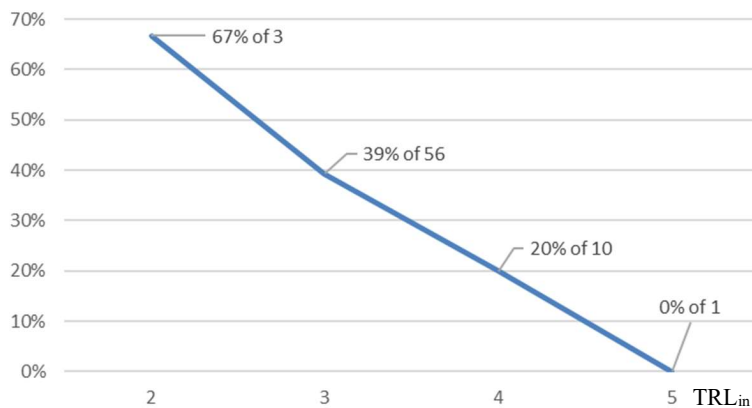

**Fig. 5.** Percentage of strategic technology development projects achieving TRL advancement of at least 1 vs. $TRL_{in}$ (e.g., 67% of three projects entering at TRL 2 advanced). Note that six of 56 projects entering with a $TRL_{in}$ of 3 advanced their TRL by two levels to TRL 5.

TRL advancement is one of the most visible success metrics for a technology maturation project. However, measuring advances solely by TRL is misleading. Each of the 70 projects made progress in maturing their technology toward the next level. The issue is that the granularity of the TRL scale is too coarse to easily capture such incremental progress. Accordingly, the Programs monitor and assess other success metrics such as technology infusion into a mission or project. Here, we define infusion as a technology being implemented in a mission/project, baselined by a mission, or incorporated into a strategic mission concept's reference design. APD's significant investment and the massive efforts and expertise that went into studying the four large-mission concepts and 10 Probes resulted in incredibly comprehensive, thorough, detailed, and compelling reports. Having any of these large-mission or Probe concepts baseline a technology into their reference design is a noteworthy accomplishment. Counting those as projected infusions reflects the importance and impact of the technology projects.

In this arena, APD's technology maturation efforts show spectacular success, with 155 infusions counted to date (summarized in Table 4 with detailed lists in Tables 5, 6, and 7). All three Programs have seen remarkable success, with technologies infused into a variety of missions and projects. These include space missions, suborbital missions and platforms (balloons, sounding rockets, and SOFIA), and ground-based observatories. Missions and projects benefiting from these technologies have naturally been primarily Astrophysics ones, but include Earth-Science, Planetary, and Heliospheric missions also.

| Decade | Space | Rocket | Balloon | Airborne | Ground | Total |
|---|---|---|---|---|---|---|
| 2010s | 8 | 14 | 8 | 2 | 24 | 56 |
| 2020s | 21 | 11 | 2 | - | 7 | 41 |
| 2030s+ | 58 | - | - | - | - | 58 |
| Total | 87 | 25 | 10 | 2 | 31 | 155 |

**Table 4.** Summary of technology infusions by decade and mission/project type. Numbers for 2010s are actuals, for 2020s include both actual and projected, and for 2030s and beyond are all projected.

| Project Type | PI | Technology | Infused Into |
|---|---|---|---|
| Space Mission | Bautz, Mark | Directly deposited optical blocking filters | OSIRIS-REx |
| Space Mission | Kilbourne, Caroline | Si-thermistor/HgTe microcalorimeter array | Hitomi |
| Space Mission | Klipstein, William | Phasemeter | GRACE-FO |
| Space Mission | Quijada, Manuel | UV coatings | GOLD and ICON |
| Space Mission | Siegmund, Oswald | APD-funded MCPs | ICON, GOLD, JUNO-UVS |
| Sounding Rocket | Greenhouse, Matthew | Next-gen microshutter arrays | FORTIS |
| Sounding Rocket | Kilbourne, Caroline | Si-thermistor/HgTe microcalorimeter array | XQC |
| Sounding Rocket | McEntaffer, Randall | X-ray reflection grating | WRXR |
| Sounding Rocket | Nikzad, Shouleh | ALD mirror coating | SISTINE |
| Sounding Rocket | Siegmund, Oswald and Vallerga, John | APD-funded MCPs | FIRE, SLICE, EUNIS, FORTIS, VeSpR, CHESS, SISTINE, DEUCE |
| Sounding Rocket | Ullom, Joel | TES microcalorimeters | Micro-X |
| Sounding Rocket | Ullom, Joel | Time-division SQUID multiplexers | Micro-X |
| Balloon | Bock, James | Antenna-coupled detectors | Spider |
| Balloon | Hu, Qing | 4.7 THz local oscillator | STO-2 |
| Balloon | Mehdi, Imran | Heterodyne detectors | STO-2 |
| Balloon | Nikzad, Shouleh | Advanced CCD detectors | FIREBall 2 |
| Balloon | Staguhn, Johannes | Far-IR large-format detectors | PIPER |
| Balloon | Ullom, Joel | Time-division SQUID multiplexers | Spider, PIPER |
| Balloon | Zmuidzinas, Jonas | TiN Kinetic Inductance Detectors (KIDs) | BLAST-TNG |
| Airborne | Mosely, Harvey | TES Bolometers | HAWC+ (SOFIA) |
| Airborne | Ullom, Joel | Time-division SQUID multiplexers | HAWC+ (SOFIA) |
| Ground-Based | Bock, James | Antenna-coupled detectors | BICEP2, BICEP3/Keck |
| Ground-Based | Nikzad, Shouleh | Delta-doped CCDs | Palomar-WaSP, ZTF |
| Ground-Based | Ninkov, Zoran | DMDs | 4.1-m SOAR telescope |
| Ground-Based | Staguhn, Johannes | TES bolometers | Bolometer camera at IRAM 30m telescope |
| Ground-Based | Ullom, Joel | Microwave SQUID multiplexers | MUSTANG2, Simons Observatory |
| Ground-Based | Ullom, Joel | OrthoMode Transducer (OMT)-coupled TES bolometers | ABS, ACTPol, AdvancedACT, ALI-CPT, MUSTANG2, SPTPol, Simons Observatory |
| Ground-Based | Ullom, Joel | TES bolometers | SCUBA2 |
| Ground-Based | Ullom, Joel | Time-division SQUID multiplexers | ABS, ACT, ACTPol, AdvancedACT, BICEP2, BICEP3/Keck, SCUBA2 |
| Ground-Based | Wollack, Edward | Feedhorn-coupled detectors | CLASS |

**Table 5.** Details of 56 actual technology infusions in 2010s by mission/project type.

| Project Type | PI | Technology | Infused Into |
|---|---|---|---|
| Space Mission | Fleming, Brian | Protected enhanced LiF mirror coatings | SPRITE CubeSat |
| Space Mission | Fleming, Brian | MCP anti-coincidence shielding | SPRITE CubeSat |
| Space Mission | Kasdin, Jeremy | Wavefront control with two deformable mirrors | Roman Space Telescope |
| Space Mission | Kasdin, Jeremy; Casement, Susan; Glassman, Tiffany; and Cash, Webster | Starshade technologies | Starshade-Roman Rendezvous Probe |
| Space Mission | Kilbourne, Caroline | Si-thermistor/HgTe microcalorimeter array | XRISM |
| Space Mission | Kilbourne, Caroline | TES Microcalorimeter arrays | ATHENA X-IFU |
| Space Mission | Kilbourne, Caroline | Time-Domain Multiplexing (TDM) | ATHENA X-IFU |
| Space Mission | Krist, John and Shaklan, Stuart | End-to-end Coronagraph models | Roman Space Telescope |
| Space Mission | Lee, Adrian | CMB detectors | LiteBIRD |
| Space Mission | Nikzad, Shouleh | Advanced CCD detectors | SPARCS CubeSat |
| Space Mission | Nikzad, Shouleh | Delta-Doped CCDs | GUCI |
| Space Mission | Quijada, Manuel | $GdF_3$ coatings | GUCI |
| Space Mission | Rauscher, Bernard | H4RG IR detectors | Roman Space Telescope |
| Space Mission | Siegmund, Oswald and Vallerga, John | APD-funded MCPs | SPRITE, ESCAPE, JUICE-UVS, EUROPA-UVS, EUVST, GLIDE, Solar Orbiter |
| Space Mission | Trauger, Jim | Hybrid Lyot Coronagraph | Roman Space Telescope |
| Sounding Rocket | Fleming, Brian | Electron-beam-lithography-ruled gratings | CHESS, DEUCE |
| Sounding Rocket | Fleming, Brian | Image Slicer | INFUSE |
| Sounding Rocket | McEntaffer, Randall | X-ray reflection gratings | OGRE, TREXS |
| Sounding Rocket | Nikzad, Shouleh | Superlattice-doped detector | SHIELDS |
| Sounding Rocket | Schattenburg, Mark and Heilman, Ralf | Blazed soft x-ray reflection grating | MaGIXS |
| Sounding Rocket | Siegmund, Oswald and Vallerga, John | APD-funded MCPs | INFUSE, DICE, Herschel |
| Sounding Rocket | Zhang, William | Single-crystal silicon X-ray mirrors | OGRE |
| Balloon | Hu, Qing | 4.7 THz local oscillators | GUSTO |
| Balloon | Mehdi, Imran | THz heterodyne arrays | ASTHROS |
| Ground-Based | Bock, James | Antenna-coupled detectors | BICEP Array |
| Ground-Based | Guyon, Olivier | Linear Wavefront Control | Subaru observatory |
| Ground-Based | Serabyn, Eugene | Vortex coronagraph | Palomar, Keck, Subaru observatories |
| Ground-Based | Ullom, Joel | OMT-coupled TES bolometers | CMB-S4 |
| Ground-Based | Ullom, Joel | TiN KIDs | Toltec |

**Table 6.** Details of 41 actual and projected technology infusions in 2020s by mission/project type.

| Project Type | PI | Technology | Infused Into |
|---|---|---|---|
| Space Mission | Bautz, Mark | Directly deposited optical blocking filters | Lynx |
| Space Mission | Bautz, Mark | Advanced CCD detector | AXIS Probe |
| Space Mission | Belikov, Ruslan | Multi-star Wavefront Sensing and Control | HabEx, LUVOIR, Roman Space Telescope |
| Space Mission | Belikov, Ruslan | PIAACMC | HabEx, LUVOIR |
10

| Space Mission | Bierden, Paul | MEMS deformable mirrors | HabEx, LUVOIR |
|---|---|---|---|
| Space Mission | Bock, James and O'Brient, Roger | CMB detectors | PICO |
| Space Mission | Bock, James and O'Brient, Roger | Antenna-coupled detectors | PICO |
| Space Mission | Conklin, John | Charge Management System (CMS) | US contribution to LISA |
| Space Mission | Hall, Don and Bottom, Michael | Avalanche Photodiode HgCdTe Near-IR detectors | HabEx, LUVOIR |
| Space Mission | Fleming, Brian | MCP Anti-coincidence shielding | LUVOIR |
| Space Mission | Greenhouse, Matthew | Next-gen microshutter arrays | HabEx, LUVOIR, CETUS Probe |
| Space Mission | Guyon, Olivier | Linear wavefront control | HabEx, LUVOIR |
| Space Mission | Guyon, Olivier | Predictive wavefront control | HabEx, LUVOIR |
| Space Mission | Guyon, Olivier | Sensor fusion | HabEx, LUVOIR |
| Space Mission | Livas, Jeffrey | Telescope | US contribution to LISA |
| Space Mission | Mauskopf, Philip | Low-power FPGA-based readout electronics for superconducting detector arrays | PICO, Origins, GEP, CDIM Probe |
| Space Mission | Nikzad, Shouleh | Delta-doped Electron-Multiplying CCDs | HabEx |
| Space Mission | Nikzad, Shouleh | Delta-doped CMOS detector arrays | LUVOIR |
| Space Mission | Schattenburg, Mark | Critical-Angle-Transmission X-ray gratings | Lynx |
| Space Mission | Schattenburg, Mark | Thermal oxide coating-stress compensation | Lynx, AXIS Probe, TAP |
| Space Mission | Serabyn, Eugene | Vortex Coronagraph | HabEx, LUVOIR |
| Space Mission | Siegmund, Oswald and Vallerga, John | APD-funded MCPs | CETUS, HabEx, LUVOIR |
| Space Mission | Siegmund, Oswald and Vallerga, John | Cross-strip MCP detector systems | HabEx, LUVOIR, CETUS |
| Space Mission | Soummer, Rémi | Apodized Pupil Lyot Coronagraph | LUVOIR |
| Space Mission | Staguhn, Johannes | Superconducting kilo-pixel far-IR detectors | Origins |
| Space Mission | Stahl, H. Philip | Predictive Thermal Control | Pathfinder for HabEx zonal thermal control |
| Space Mission | Tuttle, James | Continuous Adiabatic Demagnetization Refrigerator (CADR) | Lynx, Origins, PICO, GEP |
| Space Mission | Ullom, Joel | Microwave SQUID multiplexers | Lynx, Origins |
| Space Mission | Ullom, Joel | Time-division SQUID multiplexers | PICO |
| Space Mission | Yu, Anthony | Laser technology | US contribution to LISA |
| Space Mission | Zhang, William | Single-crystal-silicon X-ray mirrors | Lynx, AXIS, TAP |
| Space Mission | Ziemer, John | Micro-Newton thrusters | HabEx fine pointing and jitter suppression |

**Table 7.** Details of 58 projected technology infusions in 2030s and beyond.

While the breadth and depth of the above lists demonstrate the broad impact of APD's technology development investment, the main intention has always been to enable strategic Astrophysics missions. The following subsections capture some of the most significant achievements in this regard.

### 3.1 Technologies for the Roman Space Telescope's Coronagraph Instrument

The 2010 Decadal Survey, "New Worlds, New Horizons in Astronomy and Astrophysics" [14] recommended the creation of a New Worlds Technology Development Program "*…to lay the technical and scientific foundation for a future mission to study nearby Earth-like planets.*" Toward this end, APD established the Technology Development for Exoplanet Missions (TDEM) competed grant program that later became part of the SAT program. The first TDEM awards explored a wide variety of high-contrast coronagraph technologies capable of characterizing terrestrial exoplanets in the habitable zone of Sun-like stars.

In addition, as the astrophysics community and NASA studied the Decadal Survey's recommendation to develop a flagship-scale wide-field infrared survey mission (that eventually became the Roman Space Telescope), an opportunity emerged for a larger aperture telescope, making the addition of a high-contrast coronagraph instrument (CGI) technology demonstrator on that mission compelling [18]. A directed pre-phase-A activity was created to advance the technologies needed for CGI. This included coronagraph masks and architecture, in particular Hybrid Lyot [19], Shaped Pupil [20], Phase-Induced Amplitude Apodization [21] coronagraphs, low-order wavefront sensing and control [22], and electron-multiplying CCD detectors [23]. The projects maturing these technologies through TDEM-funded activities were folded into the technology activity and a technology development plan created a series of nine milestones to bring these technologies to TRL 5. The last milestone achieved in 2016 was a demonstration of coronagraph performance in a dynamic environment [24]. The Roman mission is in Phase C as of 2020, enabled by the infusion of technology developed first in a competed program, and later by a focused directed activity.

### 3.2 Technologies for ATHENA

In 2014, ESA selected ATHENA, an X-ray observatory as the second large (L-class) mission in its Cosmic Vision 2015-2025 plan under the "Hot and Energetic Universe" science theme. The mission is expected to be "adopted" in 2021. ATHENA will map the hot gas structures in the universe and reveal supermassive black holes in galaxies far enough away to peer into the early universe. To achieve this, ATHENA will carry two instruments, the X-IFU and the Wide-Field Imager (WFI). APD-funded efforts have been developing TES microcalorimeters and a TDM readout. These are currently at TRL 4 and are working toward demonstrating TRL 6 before ATHENA's Mission Adoption Review (MAR).

### 3.3 Technologies for LUVOIR

The LUVOIR mission concept was an APD-commissioned flagship study. Two large segmented architectures were considered, an on-axis design with a 15-m primary and an off-axis design with an 8-m primary. APD is investing in a wide variety of LUVOIR-enabling technologies. In particular, the extreme-contrast coronagraph instrument requires several key technologies, including an ultra-stable telescope/observatory system that must maintain wavefront stability at the tens of picometers level over 10 minutes, and segmented-aperture coronagraphy capable of imaging Earth-like planets orbiting Sun-like stars.

The APD-funded, industry-led SLSTD studies have developed models of large segmented telescopes to study their thermal and mechanical stability. In Phase 2, the teams are developing relevant technologies such as mechanical isolation and capacitive edge sensors for mirror segments. Several SAT awards are aimed at demonstrating coronagraph architectures that can work with segmented telescopes, such as the Apodized Pupil Lyot Coronagraph award. In addition, a directed modeling effort called the Segmented Coronagraph Design and Analysis study has been evaluating the link between the telescope segment phasing and the performance of different coronagraph architectures in the presence of telescope-segment phasing errors.

### 3.4 Technologies for HabEx

The HabEx concept was another APD-commissioned flagship mission study based around a telescope with a 4-m monolith primary mirror for astrophysics in the visible, near-IR, and UV. The system is designed to enable the detection and characterization of terrestrial exoplanets and uses both a coronagraph and a formation-flying starshade to perform starlight suppression. Like LUVOIR, HabEx requires extreme stability for its coronagraph, and the study baselined replacing reaction wheels with low-noise microthruster technology originally developed for the LISA Pathfinder mission. In addition, starshade technologies are being advanced by the ExEP's directed Starshade Technology Development Activity, known as S5, to meet performance requirements for either a Roman Space Telescope rendezvous or HabEx mission, both pending 2020 Decadal Survey recommendations.

### 3.5 Technologies for Origins

Origins is another of the four large mission concept studies, with a 5.9-meter primary mirror cryo-cooled to 4.5 K. Origins is planned to carry three instruments. First, a mid-IR instrument (MISC-T) to measure 2.8-20-μm spectra from transiting exoplanets, searching for bio-signatures. Next, the Far-IR Polarimeter (FIP) to carry out broadband imaging from 50 to 250 μm across thousands of square degrees. Finally, the Origins Survey Spectrometer (OSS) will carry out wide-area and deep spectroscopic surveys from 25 to 588 μm. A fourth instrument, the Heterodyne Receiver for Origins (HERO) was suggested as an up-scope option. The guiding principle for the Origins STDT was to minimize its complexity (and thus cost and risk) by building on the Spitzer architecture and minimizing deployments.

APD technology investments enabling Origins include a superconducting kilo-pixel far-IR detector architecture, a CADR sub-K cooling system, microwave SQUID multiplexers, and ultra-stable mid-infrared detectors for transit spectroscopy.

### 3.6 Technologies for Lynx

The Lynx concept is the fourth flagship-class mission study, using X-ray measurements to observe directly the dawn of super-massive black holes, look at drivers of galaxy formation, trace stellar activity and its effects on planet habitability, and study the endpoints of stellar evolution. The Lynx DRM includes a 3-meter outer diameter assembly of nearly 40,000 thin grazing-incidence mirror segments, providing a total effective area greater than 2 $m^2$ at 1 keV, with an on-axis angular resolution of 0.5" half-power diameter. Sub-arcsec resolution extends out to 10 arcmin off axis. The DRM includes three instruments. First, the X-ray microcalorimeter, providing energy resolution better than 3 eV between 200 eV and 7 keV, imaging with 1" pixels over a 5'×5' field of view (FOV). The array design includes three sub-arrays optimized for sub-arcsec imaging, 0.3-eV energy resolution, and a larger, 20'×20' FOV. The High-Definition X-ray Imager (HDXI) comprises an array of silicon sensors with 0.3" pixels extending beyond a 20'×20' FOV with ~100-eV resolution from 100 eV to 10 keV. Finally, the X-ray Grating Spectrometer will provide R>5000 resolving power with an effective area greater than 4000 $cm^2$.

APD investments enabling Lynx include single-crystal-silicon X-ray mirrors and thermal-oxide-coating stress compensation, CAT X-ray gratings, directly deposited optical blocking filters, and the previously mentioned microwave SQUID multiplexers and CADR.

### 3.7 Technologies for Probe Missions

While the 10 Probe studies funded by APD all attempted to minimize their dependence on technology development, several APD-funded technology investments have already been baselined by Probe study reference designs, including:

- For the Advanced X-ray Imaging Satellite (AXIS) Probe, an advanced CCD detector and the same grazing-incidence X-ray mirrors and thermal-oxide coatings as baselined by Lynx.
- For the Transient Astrophysics Probe (TAP), the same grazing-incidence X-ray mirrors and thermal-oxide coatings.
- For the Galaxy Evolution Probe (GEP), the same CADR as baselined by Origins and Lynx and the low-power FPGA-based readout electronics baselined by Origins.
- For PICO, CMB detectors and antenna-coupled detectors, the same CADR and low-power FPGA-based readout electronics baselined by Origins, and a time-division SQUID multiplexer.
- For the Cosmic Evolution Through UV Spectroscopy (CETUS) Probe, the same next-generation microshutter arrays as baselined by HabEx and LUVOIR, MCP detectors, and cross-strip MCP detector systems.
- For the Cosmic Dawn Intensity Mapper (CDIM) Probe, the same FPGA-based readout electronics baselined by Origins.

### 3.8 Other Exoplanet Technology Investments

While HabEx and LUVOIR represent the best recent studies of exoplanet missions, other technologies are needed to address the question "Are We Alone?" As an example, measurement of the mass of directly imaged exoplanets is critical for interpreting spectra to assess habitability if bio-signature gases are discovered in their atmospheres. The mass of a planet can be inferred by measurements of the reflex motion of the star that it orbits, either through micro-arcsec astrometry or centimeter-per-second EPRV via Doppler shifts of stellar lines. One SAT award demonstrated TRL 4 for a diffractive pupil technique to enable astrometric measurements at a level of $5.75 \times 10^{-5}$ λ/D, equivalent to 1.4 micro-arcsec on a 4-meter telescope [25]. An APRA award to the same group is continuing vacuum demonstrations to mature technology toward TRL 5.

### 3.9 Other Benefits of APD Technology Investments

In addition to advancing technologies, a majority of the PIs were able to leverage their projects and bring in additional funding, generate collaborations or other interest in their technologies, etc. This includes:

- Institutional internal funding (e.g., GSFC and MSFC Internal Research and Developments, IRADs)
- Parallel APRA and/or IRAD funding for related efforts
- Co-funding:
  - From NASA's Space Technology Mission Directorate (STMD)
  - From the Defense Advanced Research Projects Agency (DARPA)
  - From the National Science Foundation (NSF)

- Fellowships and undergraduate research funding:
  - Roman Technology Fellowship in Astrophysics
  - Leon Van Speybroeck Fellowships in X-ray Optics from SAO
  - National Research Council fellowship
  - NSF post-doctoral fellowship
  - NASA Postdoctoral Program fellowship
  - University funding for undergraduate research
- Setting up collaborations with researchers at other institutions on proposals and new programs
- Generating industry interest in their technologies
- Getting inducted as a fellow in the National Academy of Inventors

Most PIs reported hiring undergraduate and graduate students and/or post-doctoral fellows for their projects. This helped train the future astrophysics workforce and contributed to the wider US technological workforce. In total, over 100 students and post-docs participated in Astrophysics technology development projects. Many graduated with a PhD, were accepted into graduate programs, and/or obtained full-time research positions. One started a nano-fabrication business supporting the lab from which he graduated. Many post-docs proceeded to positions at other institutions or high-tech companies.

## 4. CURRENT TECHNOLOGY GAP PRIORITIES

The most recent APD technology gap prioritization was done in 2019, with technologists from all three Programs participating in each of the three separate efforts. Before examining the assigned priority tiers, a word on how gaps are prioritized, as they are meant to reflect the timely needs of the APD. The three Programs prioritize technology gaps by assigning each a score from 0 to 4 on four metrics and then using a weighted sum.

- **Strategic Alignment:** How well does closing the gap align with astrophysics science and programmatic priorities?
- **Benefits and Impacts:** How much impact does the technology have on astrophysics in applicable mission(s)? To what degree would technologies closing the gap enable and/or enhance achievable science objectives, reduce cost, and/or reduce mission risks?
- **Urgency:** How large a schedule margin do we have for closing the gap before the technologies need to be at TRL 6?
- **Scope of Applicability:** How crosscutting would closing this gap be? How many Astrophysics programs and/or mission concepts could it benefit, with an emphasis on strategic missions?

All three Programs use the same criteria, weights, and scoring guidelines. Once the three lists are completed, they're merged into a single Astrophysics technology gap priority list, broken into five descending priority tiers, with all gaps within a tier considered to have equal priority. The results of the 2019 prioritization process shown below were published in the 2019 Astrophysics Biennial Technology Report (ABTR) [26].

**Tier 1:**
- Angular Resolution (UV/Vis/Near-IR)
- Coronagraph Contrast
- Coronagraph Contrast Stability
- Cryogenic Readouts for Large-Format Far-IR Detectors
- Fast, Low-Noise, Megapixel X-Ray Imaging Arrays with Moderate Spectral Resolution
- High-Efficiency X-Ray Grating Arrays for High-Resolution Spectroscopy
- High-Resolution, Large-Area, Lightweight X-Ray Optics
- Large-Format, High-Resolution, UV/Vis Focal Plane Arrays
- Large-Format, High-Spectral-Resolution, Small-Pixel X-Ray Focal-Plane Arrays
- Large-Format, Low-Noise and Ultralow-Noise Far-IR Direct Detectors
- Large-Format, Low-Noise, High-Quantum-Efficiency Far-UV Detectors
- Next-Generation, Large-Format, Object Selection Technology for Multi-Object Spectrometers for LUVOIR
- Vis/Near-IR Detection Sensitivity

**Tier 2:**

- Advanced Millimeter-Wave Focal-Plane Arrays for CMB Polarimetry
- Detection Stability in Mid-IR
- Heterodyne Far-IR Detector Arrays and Related Technologies
- High-Efficiency Object Selection Technology for UV Multi-Object Spectrometers
- High-Performance Spectral Dispersion Component/Device
- High-Reflectivity Broadband Far-UV-to-Near-IR Mirror Coatings
- High-Throughput Bandpass Selection for UV/Vis
- Large-Format Object Selection Technology for Multi-Object Spectrometers for HabEx
- Starshade Deployment and Shape Stability
- Starshade Starlight Suppression and Model Validation
- Stellar Reflex Motion Sensitivity – Astrometry

**Tier 3:**

- Advanced Cryocoolers
- High-Performance, Sub-Kelvin Coolers
- Large Cryogenic Optics for the Mid-IR to Far-IR
- Long-Wavelength-Blocking Filters for X-Ray Micro-Calorimeters
- Low-Noise, High-Quantum-Efficiency UV Detectors
- Low-Stress, Highly Stable X-Ray Reflective Coatings
- Photon-Counting, Large-Format UV Detectors
- Polarization-Preserving Millimeter-Wave Optical Elements
- UV Coatings
- UV Detection Sensitivity
- UV/Vis/Near-IR Tunable Narrow-Band Imaging Capability
- Warm Readout Electronics for Large-Format Far-IR Detectors

**Tier 4:**

- Compact, Integrated Spectrometers for 100 to 1000 μm
- Optical-Blocking Filters
- Rapid Readout Electronics for X-Ray Detectors
- Short-Wave UV Coatings
- Warm Readout Electronics for Large-Format Far-IR Detectors

**Tier 5:**

- Advancement of X-Ray Polarimeter Sensitivity
- Far-IR Spatio-Spectral Interferometry
- High-Precision Low-Frequency Radio Spectrometers and Interferometers
- Mid-IR Coronagraph Contrast
- Ultra-High-Resolution Focusing X-Ray Observatory Telescope
- Very-Wide-Field Focusing Instrument for Time-Domain X-Ray Astronomy
- Wide-Bandwidth, High-Spectral-Dynamic-Range Receiving System for Low-Radio-Frequency Observations on the Lunar Far Side

The next prioritization exercise is planned for mid-2021, following the release of 2020 Decadal Survey recommendations, and updated strategic guidance from APD. The updated priority list is planned to be published in October 2021.

## 5. SUMMARY

We presented an analysis of APD's strategic technology investments since 2009 in a wide range of topics, including optics, detectors, coatings, coronagraphs, starshades, lasers, electronics, cooling, and micro-thruster subsystems. The analysis included the distribution of projects and investment between the three Programs, organizational type of PI institutions, technology topics (showing 67% of investigations addressing the critical topics of detectors, coronagraphs, and optics), a 37% TRL advancement rate overall, and how a higher entry TRL makes TRL advancement more challenging. Given the relatively modest investment of $209 million in 70 projects, we believe the demonstrated return on that investment has been profound. This includes 155 technology infusions, 32 TRL step advances, over 100 students and postdoctoral fellows trained, and a long list of collaborations and other contributions. We listed the current portfolio of technology investigations across the three APD Programs, and the most recent Astrophysics technology gap priorities.

The 2020 Astronomy and Astrophysics Decadal Survey report is expected to be published in the spring of 2021, which may well modify APD's strategic guidance to the three Programs, and thus technology-gap priorities. APD and the PCOS, COR, and ExE Programs will endeavor to identify and advance the key technologies for the highest recommended missions and projects over the coming decade.

We encourage all PIs receiving APD technology-development funding involve in their projects as many postdoctoral fellows, graduate and undergraduate students, and interns as feasible. This will continue training our astrophysics workforce and expand inclusion of all genders and under-represented minorities in the Science, Technology, Engineering, and Math (STEM) fields, promoting a brighter future for NASA Astrophysics and its inspiring work.